\documentclass[prb,aps,amssymb,showpacs,twocolumn,amsmath,floatfix]{revtex4}
\usepackage{tabularx}
\usepackage{bm}
\usepackage{euscript}
\usepackage{epsfig,psfrag,subfigure}
\usepackage{graphicx}
\usepackage{color}
\usepackage{amsfonts}
\usepackage{exscale}
\usepackage{amsbsy}
\usepackage{hyperref}
\input{epsf}

\def\be{\begin{equation}}
\def\ee{\end{equation}}
\def\ba{\begin{eqnarray}}
\def\ea{\end{eqnarray}}

\begin{document}

\title{Superconductivity from repulsive interactions in the two dimensional electron gas  }
\author{S. Raghu$^{1,2}$, and S. A. Kivelson$^1$}
\affiliation{$^1$Department of Physics, Stanford University, Stanford, CA 94305}
\affiliation{$^2$Department of Physics and Astronomy, Rice University, Houston, TX 77005}
\date{\today}

\begin{abstract}
We present a well-controlled perturbative renormalization group (RG) treatment  of superconductivity from short-ranged repulsive 
interactions in a variety of model two dimensional electronic systems.  Our analysis applies in the limit where the repulsive interactions between the electrons are small compared to their kinetic energy.   
\end{abstract}

\pacs{74.20.-z, 74.20.Mn, 74.20.Rp, 74.72.-h}

\maketitle

\section{Introduction}

In a variety of recently discovered materials, superconductivity  apparently arises directly  from the electron correlations themselves.
However, these materials are complex, and many material specific details
can play a role in the mechanism of superconductivity. The problem is greatly simplified in the weak coupling limit, where we recently showed \cite{Raghu2010}  that an asymptotically exact treatment of the problem is possible, valid in cases in which the superconducting state emerges at low temperatures from a well-formed Fermi liquid.  Nonetheless, even under these circumstances, the character of the superconducting state and the  transition temperature depend in a complicated way on details of the band-structure, both near and far from the Fermi surface.

 To the extent that there are basic principles at work underlying the mechanism of unconventional superconductivity, it would be a great advance to find  simple model systems which exhibit such behavior.  Here, we consider the possibility of unconventional superconductivity in some model systems with particularly simple electronic structures, where controlled theory is possible, and where, conceivably, 
  experimental tests of the theory are feasible.  Specifically, we consider circumstances in which superconductivity may occur in a  two dimensional electron gas (2DEG) in a high mobility 
 heterostructure.    
  Here,  
  due to the the stiffness of the lattice and the limited phase space for electron-phonon scattering, electron-phonon coupling is probably negligible, and
  the single-particle dynamics can be treated accurately within a rotationally invariant effective mass approximation. Moreover, the strength of the correlations can, to a large extent, be tuned by varying the electron density.

The possibility of an electronic pairing mechanism in systems with rotational invariance was first put forth in a seminal paper by Kohn and Luttinger\cite{Kohn1965}.  
Although $U$,  the bare interactions among electrons are repulsive, there are {\it effective} attractive interactions that arise at $\mathcal{O}(U^2)$.  Kohn and Luttinger focussed on the portion of the effective attractions   
associated with the non-analyticities in $\chi(q)$, the particle-hole susceptibility, at  momentum $q= 2k_F$ which reflect  the sharpness of the Fermi surface at zero temperature.
More generally, what is required for this mechanism to work is strong $q$ dependence of $\chi(q)$ for $q \leq 2k_F$.  Indeed, the Kohn Luttinger instability of a 3 dimensional rotationally invariant system results in the formation of a p-wave superconducting ground state due to the peak in 
$\chi(q)$ near $q=0$.  While this result is valid only in the weak-coupling regime where $U \ll E_F$, it is widely believed that the p-wave ground state obtained this way is  adiabatically connected to the more realistic (and more strongly correlated) example of Helium-3\cite{Leggett1975}.

However, the Kohn-Luttinger effect is exponentially weaker in a rotationally invariant 2DEG\cite{Galitski2003}, due to 
the fact 
$\chi(q)$ is 
 independent of momentum for 
momenta $q \le 2k_F$.  
It was later shown 
that at $\mathcal{O}(U^3)$, the 2DEG does exhibit 
a pairing instability\cite{Chubukov1992}.  
Still, at least in weak-coupling, electronically mediated superconductivity in the 2DEG is 
negligible.  

In this paper, we 
show  that by perturbing the 2DEG, it is possible to 
  significantly enhance the superconducting 
  transition temperature 
  by engineering circumstances in which instabilities 
  arise at $\mathcal{O}(U^2)$ in perturbation theory.  
  We present asymptotically exact\cite{Raghu2010} weak coupling solutions of the superconducting instability in several systems that are variants of the simplest, rotationally invariant 2DEG.  
  As a first example, we show that 
  partially spin-polarizing the 2DEG 
  produces 
  a non-unitary $p+ip$ superconductor.  
Y. Kagan and A. Chubukov have previously addressed this problem using 
  an expansion in powers of the electron concentration
\cite{Kagan1989}, and their result reduces to ours in the weak coupling limit.   
As a second example, we consider 
 the 2DEG in a semiconductor heterostructure quantum well with two populated subbands.  We show that this system can possess both p-wave and d-wave ground states and present the phase diagram of this system.  
 
This paper is organized as follows.  In the next section, we review the method developed in Ref. \cite{Raghu2010} and discuss its straightforward generalization needed for the present context.  In Section III, the effect of partially polarizing the 2DEG is studied.  In Section 
IV, we consider 
 two subbands in a 2DEG quantum well.  
Technical details of the various calculations are presented in the Appendix.  In a forthcoming paper\cite{RaghuII} we will consider a variety of slightly more complex situations pertinent to particular semiconductor heterostructures.

\section{Perturbative renormalization group treatment of superconductvity}
In this section, we review the prescription of Ref. \cite{Raghu2010} and discuss 
its generalization 
to the present context.  
We integrate out high energy modes in two steps.
In the first step, we integrate out all 
modes outside a narrow range of energies $\Omega$ about the Fermi energy.  $\Omega$ is not a physical energy in the problem, but rather a calculational device.  It is chosen large enough so that the interactions can be treated perturbatively but small enough that it can be set to zero in all non-singular expressions without causing significant error, {\it i.e.} it is chosen to satisfy the inequalities $\rho U^2 \gg \Omega \gg \mu\exp\{-[1/\rho U] \}$, 
where $\rho$ is the density of states at the Fermi energy and $\mu$ is the Fermi energy.  The effective interactions generated in the process then serve as the ``bare'' interactions in a second step, in which the remaining problem is solved using the perturbative renormalization group procedure of Shankar and Polchinski\cite{Shankar1994,Polchinski1992}.  $T_c$ is, up to an unknown multiplicative constant, given by the energy scale, $T^*$, at which a relevant interaction grows to be of order 1.  It was shown by careful analysis of perturbative expressions up to 4th order in the interaction strength that the resulting expression for $T^*$ is independent of 
$\Omega$.  

The analysis of Ref. \cite{Raghu2010} leads to the following prescription for computing the leading order asymptotic behavior of $T_c$ for weak interactions:  First, compute the effective interaction in the Cooper channel at energy scale $\Omega$, $\Gamma^{(a)}(\hat k,\hat k^\prime)$, to second order in the interactions.  Here, $\hat k$ and $\hat k^\prime$ denote points on the Fermi surface, and  $\Gamma$ is the vertex for scattering a pair of electrons with momenta $\hat k$ and $-\hat k$ to states with momenta  $\hat k^\prime$ and $-\hat k^\prime$, where if there are multiple band indices,
the subband index is implicitly determined depending on whether the momenta are on one Fermi surface or the other, and where there is a different matrix depending on whether the electron pair forms a spin singlet ($\Gamma^{(s)}$) or a spin triplet ($\Gamma^{(t)}$).  We then construct the related dimensionless matrix
\be
g^{(a)}_{\hat k,\hat k^\prime} \equiv  
\rho \sqrt{\bar v/v(\hat k)}\Gamma^{(a)}(\hat k,\hat k^\prime) \sqrt{\bar v/v(\hat k^\prime)},
\ee
where $v(\hat k)$ is the magnitude of the Fermi velocity on the Fermi surface of the  corresponding subband, and $\rho$ is the total density of states 
at the Fermi energy.
   Manifestly, $g$ is a real, symmetric, hence Hermitian matrix, so it has a complete set of eigenstates and eigenvalues, 
\be
\sum_{\hat k^\prime}g^{(a)}_{\hat k,\hat k^\prime} \phi_{\hat k^\prime}^{(a,m)} =\lambda^{(a,m)}\phi_{\hat k}^{(a,m)}.
\ee
Among all the possible solutions, we identify the most negative eigenvalue, 
\be
\lambda \equiv {\rm Min}\left[  \lambda^{(a,m)}\right], \lambda < 0
\ee
Then,
\be
T_c \sim \mu \exp[-1/\vert \lambda \vert ].
\ee

\section{Partially polarized Fermi surface}
As a first example, we consider 
a 
partially spin polarized 2DEG
with short-ranged repulsive interactions:
\begin{eqnarray}
H&=&H_0 + H_1 \nonumber \\
H_0 &=& \sum_{\sigma} \int \frac{d^2 k}{(2 \pi)^2} E_{\sigma, \sigma'}(\bm k) \psi^{\dagger}_{\sigma}(\bm k) \psi_{\sigma'}(\bm k) \nonumber \\
H_1 &=&U  \int   \frac{d^2 k_1 d^2 k_2 d^2 k_3}{(2 \pi)^6} 
\psi^{\dagger}_{\uparrow}(\bm k_1) \psi^{\dagger}_{ \downarrow}(\bm k_2) \psi_{ \downarrow} (\bm k_3) \psi_{\uparrow}(\bm k_4) \nonumber \\
\end{eqnarray}
where $ \bm k_4 = \bm k_1 + \bm k_2 - \bm k_3$, 
\begin{equation}
E_{\sigma, \sigma'} = \epsilon_{\bm k} \delta_{\sigma, \sigma'} + \bm h \cdot \bm \tau_{\sigma \sigma'}, 
\end{equation}
and $\bm  h$ is a mean-field 
 that 
renders the ground state 
spin-polarized.  Such a partially-polarized system can occur in a narrow-well semiconductor heterostructure in the presence of a parallel magnetic field (in which case ${\bf h} = g \mu_B {\bf H}_{\parallel}$), or in a ferromagnetic phase with spontaneously broken symmetry, such as 
probably occurs in the Hubbard model 
away from half-filling in the strong-coupling limit $U \gg t$\cite{Pilati2010,Liinprep}.  (However, in the latter case, it requires something of an intuitive leap to treat the residual interactions beyond those that produce the mean-field ${\bf h}$ as ``weak.'')

Since the Fermi surfaces are spin-polarized, singlet pairing 
is suppressed, 
so the  leading superconducting instability will therefore be in the spin triplet 
channel.  We first consider the limit in which there is no spin-orbit coupling, in which case, the two particle scattering amplitude is a separate function  for each spin-polarization.  As derived in the appendix, 
\begin{eqnarray}
\Gamma_{\uparrow} (\hat k, \hat q) = - U^2  \chi_{\downarrow}(\vec k - \vec q)   \nonumber \\
\Gamma_{\downarrow} (\hat k, \hat q) = -  U^2  \chi_{\uparrow}(\vec k - \vec q)  \nonumber
\end{eqnarray}
where $\chi_{\sigma}$ is the contribution of spin $\sigma$ electrons to the susceptibility.

In  the case of a rotationally invariant system with $\epsilon_{\vec k,\sigma} = k^2/2m +\sigma h$,  
$v_{f,\sigma}(\hat k) = 
k_{f,\sigma}/m$ 
 and $\rho_{\sigma} = \rho = m/2 \pi$ is independent of the spin-polarization.  Therefore the matrix $g_{\hat k, \hat q}$ defined in the previous section is 
\begin{equation}
g^{\sigma}_{\hat k, \hat q} = \rho \Gamma_{\sigma} (\hat k, \hat k')
\end{equation}
The particle-hole susceptibility for this system has the following well-known 
form (see the appendix):
\begin{equation}
\label{chi2d}
\chi_{\sigma}(\vec q) = \frac{\rho}{2 } \left[ 1 - \frac{{\rm Re} \sqrt{q^2 - (2 k_{F \sigma})^2}}{q} \right]
\end{equation}
Thus, $\chi_{\sigma}(q)$ is  a constant for $q < 2k_{F \sigma}$, has a derivative discontinuity at $q = 2 k_{F \sigma}$,  and vanishes as $1/q^2$ when $ q >> 2k_{F \sigma}$. 

The rotational invariance of the problem 
implies that the triplet eigenfunctions are of the form 
\begin{equation}
\psi^{t,m}_{\sigma} (\hat k) = \psi(k_{F \sigma}) \cos{ \left( m  \theta_{\hat k} \right)}
\end{equation}
where $m$ is an odd integer.  
The eigenvalue problem for this system  therefore 
reduces to the integral expressions:
\begin{eqnarray}
 \lambda^{(m, \uparrow)}&=& - \rho U^2 \int \frac{d \theta}{2 \pi} \chi_{\downarrow}\left( 2k_{F \uparrow} \left| \sin{\left(\theta/2 \right) } \right| \right) \cos{ \left( m  \theta \right)} \nonumber \\
 \lambda^{(m, \downarrow)}  &=& - \rho U^2 \int \frac{d \theta}{2 \pi} \chi_{\uparrow}\left( 2k_{F \downarrow} \left| \sin{\left(\theta/2 \right) } \right| \right) \cos{ \left( m  \theta \right)} \nonumber \\ 
  \end{eqnarray}
where $\theta$ is the angle relative 
between $\hat k$ and $\hat q$.

Without loss of generality, we suppose that $k_{F \downarrow} < k_{F \uparrow}$.  
For any $\hat k, \hat q$ on the smaller (spin-down) Fermi surface, $\hat k - \hat q < 2k_{F \downarrow} < 2k_{F \uparrow}$ 
so the effective interaction, 
$\sim \chi_{\uparrow}(\hat k - \hat q)$, 
is a constant.  Therefore, it follows that $\lambda_{m \downarrow} = 0$ for all $m$ or in other words, the smaller Fermi surface 
has no superconducting instability to $\mathcal O(U^2)$. (
Presumably, $\lambda_{m,\downarrow} \sim  \mathcal O(U^3)$.   )

 Conversely,  the effective interaction between electrons on the larger (spin-up) Fermi surface is $\sim \chi_{\downarrow}(\hat k - \hat q)$, which {\it does} depend on the relative 
 position of the incoming and outgoing electrons on the Fermi surface.  
 Using Eq. \ref{chi2d}, the 
 above expression for the eigenvalue on the larger Fermi surface 
 becomes 
 \begin{equation}
 \lambda_{m \uparrow}(\eta) = \frac{\rho^2 U^2}{\pi} \int_{\theta_c}^{\pi} d \theta \frac{ \sqrt{ \sin^2{\frac{\theta}{2} } - \eta^2} }{\sin{\frac{\theta}{2}} } \cos{\left( m \theta \right) }
 \end{equation}
where $\eta =  \left( k_{F \downarrow}/k_{F \uparrow} \right)$, $0 \leq \eta \leq 1$, and $\theta_c = 2 \sin^{-1}{\eta}$.  
As can be seen from the equation above, $\lambda_{m \uparrow}(0) = \lambda_{m \uparrow}(1) = 0$.  That is, in the limit where the Fermi surface is either completely polarized, or completely unpolarized, there is no superconducting instability to $\mathcal{O}(U^2)$.  However, for intermediate values of the polarization, the integral above yields
\begin{equation}
\lambda_{1 \uparrow}(\eta) = - \rho^2 U^2 \eta \left( 1 - \eta \right)
\end{equation} 
 which is clearly negative for all intermediate values of $\eta$.  This is 
 the  main result of this section: by polarizing the Fermi surfaces in two dimensions, there is a significant enhancement of p-wave superconductivity.  The optimal pairing strength occurs when $\eta = 1/2$, so that 
 \begin{equation}
{\rm Max}\left[\lambda_{m \uparrow}(\eta) \right] =  \lambda_{1 \uparrow} (\eta = 0.5) = - \frac{ (\rho U)^2}{4}
 \end{equation}
( Note that Eq. 7 of Ref.  \cite{Kagan1989} reduces to this result in the limit of weak interaction.  )
For completeness, we quote the next leading eigenvalue, which corresponds to the f-wave (i.e. $m=3$) solution:
\begin{equation}
\lambda_{3 \uparrow}(\eta) = - \rho^2 U^2 \eta \left[ 1 - \eta\left( 3 - 4 \eta^2 + 2 \eta^4 \right) \right]
\end{equation} 
which is not symmetric about the point $\eta = 0.5$.  \\

Weak but non-vanishing spin orbit coupling will generically change this situation, since superconductivity will be induced in the minority fluid by the proximity effect as soon as the majority fluid becomes superconducting.  In 2D, this induced superconductivity will generally track the fundamental order parameter.

\section{Two subbands in a 2DEG}
\label{2subband}
In this section, we consider the case of a 2DEG in a semiconductor heterostructure having two subbands, 
with Hamiltonian:
\begin{eqnarray}
H&=&H_0 + H_1 \nonumber \\
H_0 &=& \sum_{a = 1,2} \sum_{\sigma} \int \frac{d^2 k}{(2 \pi)^2} \epsilon_{\bm k, a} \psi^{\dagger}_{a, \sigma}(\bm k) \psi_{a, \sigma}(\bm k) \nonumber \\
H_1 &=& \sum_{a.. d} \sum_{\sigma, \sigma'} V^{\sigma, \sigma'}_{ab,cd} \int   \frac{d^2 k_1 d^2 k_2 d^2 k_3}{(2 \pi)^6} [ \nonumber \\
&&\psi^{\dagger}_{a, \sigma}(\bm k_1) \psi^{\dagger}_{b, \sigma'} \psi_{c, \sigma'} (\bm k_3) \psi_{d, \sigma}(\bm k_4)]
\end{eqnarray}
where $a$ is the subband index and is used to distinguish the smaller $(a = 1)$ and larger ($a=2$) Fermi surface, 
and $\epsilon_{\bm k,a} = k^2/2m + \delta_a$ with $\delta_1 = 0$ and $\delta_2 > 0$.
The interactions are assumed to be short-ranged, consisting of an intra-band repulsion $U$, an inter-band repulsion $V$, and an inter-band pair-hopping amplitude $J$.  The interaction matrix 
in the basis 
$(1 \sigma 1 \sigma', 1 \sigma 2 \sigma', 2 \sigma 1 \sigma', 2 \sigma 2 \sigma')$ 
is thus
\begin{equation}
V^{\sigma , \sigma'}_{ab, cd} = 
\left( \begin{array}{cccc}
U_{\sigma \sigma'} & 0 & 0 & J_{\sigma \sigma'} \\
0 & 0 & V_{\sigma \sigma'} & 0 \\
0 & V_{\sigma \sigma'} & 0 & 0 \\
J_{\sigma \sigma'} & 0 & 0 & U_{ \sigma \sigma'}
\end{array} \right)
\end{equation}
where
\begin{eqnarray}
U_{\sigma \sigma'} &=& U\left( 1 - \delta_{\sigma \sigma'} \right) \nonumber \\
J_{\sigma \sigma'} &=& J\left( 1 - \delta_{\sigma \sigma'} \right) \nonumber \\
V_{\sigma \sigma'} &=& V
\end{eqnarray}

 As before, 
 rotational invariance enables us to label the eigenstates  by the eigenvalue of the rotation operator, 
  \begin{equation}
 \phi^{(m)}_{\hat k} = \phi_a^{(m)} 
 \cos{\left( m \theta \right)}
 \end{equation}
 where the complex amplitude $\phi_a^{(m)}$ depends only on the subband index associated with $\hat k$, 
 $\theta$ is the angle between $\hat k$ and an arbitrarily defined $x$ axis, and $m$ must be an even integer in the singlet channel 
 and an odd integer in the triplet channel. 
 Consequently, for each integer $m$, rotational symmetry reduces the eigenvalue problem to a $2\times 2$ problem,
 \begin{equation}
\sum_{a,a^\prime} 
\tilde g^{(m)}_{a,a^\prime} 
\phi^{(m)}_{a^\prime} 
= \lambda^{(m)}\phi^{(m)}_{a} 
\end{equation}
where
\begin{equation}
\tilde g^{(m)}_{a,a^\prime} 
\equiv \int_a 
\frac {d\hat k}{2\pi}\int_{a^\prime} 
\frac  {d\hat k}{2\pi}
g^{(y)}_{\hat k,\hat k^\prime} e^{-im\theta}e^{im\theta^\prime},
\label{tildeg}
\end{equation}
where $y=s$ (singlet) for $m$ even and $y=t$ (triplet) for $m$ odd.
The most negative eigenvalue for fixed $m$ is
\begin{eqnarray}
\lambda^{(m)} =&&- \left(\frac{\tilde g^{(m)}_{1,1}+ \tilde g^{(m)}_{2,2}} 2\right) \\
&&-\sqrt{ \left(\frac { \tilde g^{(m)}_{1,1}- \tilde g^{(m)}_{2,2}} 2\right)^2+| \tilde g^{(m)}_{1,2}|^2}
\nonumber
\end{eqnarray}

We first consider the {\bf spin triplet channel} ($m$ odd) which is only a slight extension of the result obtained for a partially polarized Fermi surface.  As shown in the Appendix, for odd $m$, the effective interaction is diagonal in the subband index, and depends on $U$, $V$,  but not $J$:  
\begin{eqnarray}
 g^{(t)}_{1,1} (\hat k, \hat q) &=& - \rho U^2 \chi_{1,1} (\hat k - \hat q) - 2 \rho V^2 \chi_{2,2} (\hat k - \hat q) \nonumber\\
 g^{(t)}_{2,2} (\hat k, \hat q) &=& -\rho U^2 \chi_{2,2} (\hat k - \hat q) - 2 \rho V^2 \chi_{1,1} (\hat k - \hat q) \nonumber \\
 g^{(t)}_{1,2} (\hat k, \hat q) & = & 0 
\end{eqnarray}
where 
\begin{equation}
\chi_{a,b}( \bm k) =  \int \frac{d^2 p}{(2 \pi)^2} \frac{f(\epsilon_{\bm p + \bm k, a}) - f(\epsilon_{\bm p, b})}{\epsilon_{\bm p + \bm k, a} - \epsilon_{\bm p, b}} 
\end{equation}
is the particle-hole susceptibility generalized to the two band system.  
The intraband susceptibilities 
 are precisely the same functions used before:
\begin{equation}
\label{chi2d2}
\chi_{a,a}(\vec q) = \frac{\rho}{2 } \left[ 1 - \frac{{\rm Re} \sqrt{q^2 - (2 k_{F a})^2}}{q} \right]
\end{equation}
with the subband index playing the role that the spin played in the previous section.  Therefore, we may simply 
transcribe the results found in the previous section to the present context.  The band which forms the smaller Fermi surface ($a=1$) does not exhibit a superconducting instability to $\mathcal{O}(U^2)$. 
The larger Fermi surface exhibits a triplet 
p-wave instability with a pairing strength determined solely by $V$: 
\begin{equation}
\lambda^{(1)}(\eta) = -4 \rho^2 V^2 \eta( 1- \eta) 
\end{equation} 
where $\eta = \left( k_{F 1}/ k_{F 2} \right)$. 
 
In the {\bf spin-singlet channel}, the matrix $g$ 
has off-diagonal components:
\begin{eqnarray}
\label{singlet}
g^{(s)}_{1,1}(\hat k, \hat q) &=& \rho U_1 
- 2\rho V^2 \chi_{2,2}(\hat k - \hat q) 
 \\
g^{(s)}_{2,2}(\hat k, \hat q) &=& \rho U_2 
- 2 \rho V^2 \chi_{1,1}(\hat k - \hat q) \nonumber \\
g^{(s)}_{1,2}(\hat k, \hat q) &=& \rho U_{12} 
+ \rho VJ  \left[ \chi_{1,2}(\hat k + \hat q) + \chi_{1,2}( \hat k - \hat q) \right] \nonumber 
\end{eqnarray}
where  $U_{ab}$ are momentum independent interactions, 
\begin{eqnarray}
U_1 &\equiv &  U  + U^2P_1(\Omega) + J^2 P_2(\Omega) + U^2 \chi_{1,1}(\hat k + \hat q) 
\nonumber \\
U_2 &\equiv &  U + U^2P_2(\Omega) + J^2 P_1(\Omega)  + U^2 \chi_{2,2}(\hat k + \hat q) 
\nonumber \\ 
U_{12} &\equiv&  UJ \left[ P_1 (\Omega) + P_2(\Omega) \right]  
\end{eqnarray}
where the particle-particle susceptibility,
\begin{eqnarray}
P_a (\Omega) &\equiv & 
\int \frac{d^2 q}{(2 \pi)^2} \frac{  2 f 
(\epsilon_{\bm q,a}) -1 }{i \Omega - 2 \epsilon_{\bm q,a}} \nonumber \\
& \sim & \rho
\log \left[ \frac{E_F-\delta_a}{\Omega} \right] + \mathcal{O}(\Omega),
\end{eqnarray}
is a momentum-independent constant which diverges logarithmically at low energies.  
Despite this divergence, the second order contributions to $U_\alpha$ are unimportant, since
they do not enter the gap equation for any $m\neq 0$, 
and any s-wave solution is  already  
killed by the first order terms proportional to $U$.

For $m>0$ and even, the effective intra-band interaction depends only on the interaction $V$ and is non-zero only for the larger Fermi surface, whereas the inter-band interaction depends both on $V$ and $J$:
\begin{eqnarray}
\tilde g_{1,1}^{(m)} &= &0 \nonumber \\
\tilde g_{2,2}^{(m)} &=& -2V^2 \rho \int \frac{d \theta}{2 \pi}   \chi_{1,1}\left( 2 k_{F2} \left | \sin \left( \theta/2 \right)\right |  \right) \cos\left( m \theta \right)  \nonumber \\
\tilde g_{1,2}^{(m)} &=& 2VJ \rho \int \frac{d \theta}{2 \pi} \chi_{1,2}
\left( k_{\theta}   \right) \cos\left( m \theta \right)  \nonumber \\
k_{\theta} &=& k_{F2} \sqrt{\left(1-\eta \right)^2 + 4 \eta \sin^2 \left( \theta/2 \right)  } .
\end{eqnarray}
The explicit expression for the inter-band susceptibility $\chi_{1,2}(\bm q) $ is derived in the Appendix.  
Since the $m=0$ 
eigenvalues  are always positive, 
the dominant singlet instability 
is in the d-wave ($m=2$) 
channel.  The quantity $\tilde g_{2,2}^{(2)}$ is obtained by computing 
\begin{eqnarray}
\tilde g_{2,2}^{(2)} &=& -2V^2 \rho^2 \int_{-\pi}^{\pi} \frac{d \theta}{2 \pi} 
d \theta \frac{ {\rm Re} \sqrt{ \sin^2{\frac{\theta}{2} } - \eta^2} }{\sin{\frac{\theta}{2}} } \cos{\left( 2 \theta \right) } \nonumber \\
&=& -V^2 \rho^2 \eta \left(\eta-1 \right)\left(\eta^2+\eta-1 \right)
 \end{eqnarray}
The interband interaction $\tilde g_{1,2}^{(2)}$ is also obtained using Eq. \ref{singlet}:
\begin{equation}
\tilde g_{1,2}^{(2)} = -\frac{V J \rho^2}{2 \pi} \Phi(\eta)
\end{equation}
where, for $0 \le x < 1$, 
\begin{equation}
\Phi(x) =   \frac{\pi x^4+2 \sin^{-1} x}{x^2}  
- 2\frac {\sqrt{1-x^2}}  x
\end{equation}

This function is discussed in detail in the Appendix.  An important property of $\Phi(x)$ is that it is  a 
monotonically increasing function of $x$ for $0 \le x <1$ (see Fig. \ref{interbandchi}).  Therefore, the effective interband scattering grows with $\eta$.  
The pairing strength in the d-wave channel is obtained from these quantities via 
\begin{equation}
\lambda^{(2)}(\eta) = \frac{\tilde g^{(2)}_{2,2}}{2} - \frac{1}{2} \sqrt{\left(\tilde g^{(2)}_{2,2} \right)^2 + 4 \left( \tilde g^{(2)}_{1,2}\right)^2 }
\end{equation}
 \begin{figure*}
\includegraphics[width=5.5in]{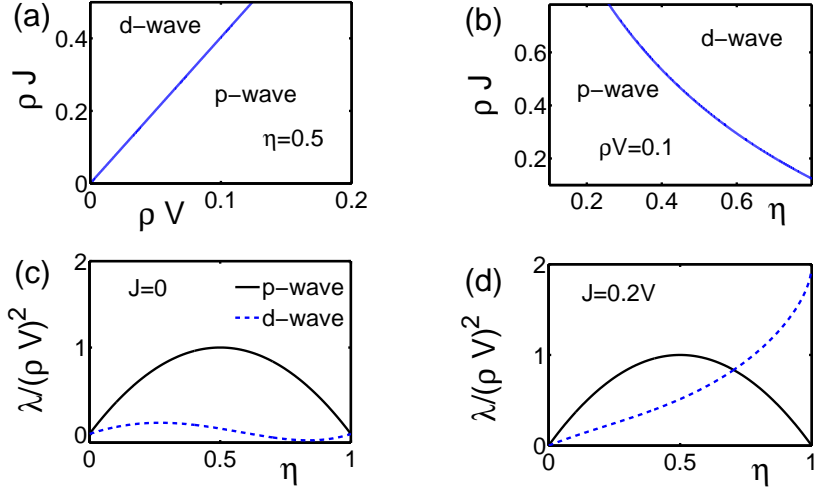}
\caption{Phase diagram of a 2DEG  having two subbands.  (a) Phase diagram for fixed $ \eta\equiv k_{F1}/k_{F2} = 0.5$ as a function of the dimensionless couplings 
$\rho V$ and $\rho J$.  $U$ does not enter the problem except in that it is responsible for the suppression of s-wave pairing.  
(b) Phase diagram for fixed 
$\rho V=0.1$ as a function of $\eta$ and 
$\rho J$.  
c) The dimensionless strength of the pairing interaction in the p-wave (solid line) and d-wave (dashed line) channels for fixed $J=0$.  d)  Same as c), but for $J=0.2V$.
}
\label{phasediag}
\end{figure*} 
Having derived closed form expressions for the p-wave and d-wave pairing strengths, we can 
construct the phase diagram, shown in Fig. \ref{phasediag}.  The phases are labeled according to the symmetry of the most negative eigenvalue, so 
 the phase boundaries 
  are the lines at which 
   $\lambda^{(2)}=\lambda^{(1)} < 0$.  Since  the d-wave and p-wave eigenvalues are both negative for all $0 \le \eta \le 1$, where one phase is stable, the other is metastable. 
 It would require different methods of analysis to completely characterize the  phase competition.  However, in weak coupling, the phase with the larger $|\lambda_m|$ has an exponentially larger $T_c$, and so gaps the entire Fermi surface at temperatures far above the putative transition temperature of the subdominant phase.  Thus, a  BCS mean-field treatment of this problem would suggest that at low temperatures, there is a direct, first order transition, or at most an exponentially narrow region of phase coexistence  between the extremal pure d-wave and p-wave phases.
 
 Since the pair-hopping term only affects spin-singlet superconductivity, and since $\vert \lambda^{(1)} \vert > \vert \tilde g_{2,2}^{(2)} \vert$, it follows that for $J=0$, the p-wave solution always remains the favored ground state, as can be seen from Fig. \ref{phasediag}(a).  However, as the interband scattering is enhanced, the d-wave pairing strength grows.  Since the interband scattering increases monotonically as a function of both $\eta$ and $J$, it is seen that for sufficiently large values of either parameter the p-wave superconductivity gives way  to a d-wave ground state.    In Figs. \ref{phasediag}(c-d), we show how the magnitude of $\lambda^{(1)}$ and $\lambda^{(2)}$ depend on $\eta$, from which one can see that $T_c$ is maximal in the p-wave channel when 
 $\eta=0.5$.  However, when $J \ne 0$, the 
 d-wave channel grows monotonically with $\eta$ and ultimately overtakes the p-wave pairing strength 
 as $\eta \rightarrow 1$.  Note, however, that $\eta$ can never 
 equal unity in this context, since it 
 is determined by the thickness of the quantum well.

\section{Discussion}
We have obtained analytical expressions for various unconventional 
superconducting ground states of a clean 2DEG in the presence of weak, short-ranged repulsive interactions.  Ultimately, to make contact with experiments involving real 2DEGs, we must take into account the Coulomb interactions.  
In the small $r_s$ limit, 
the Coulomb interactions are sufficiently well screened 
that it may be reasonable to treat them as weak and short-ranged\footnote{A diagrammatic approach to the full Coulomb problem in 3D at small $r_s$ was explored in Ref. \cite{Chubukov1989}, but there are many subtleties which make this hard to extend}.  We thus imagine we can relate the physical problem to a problem with short-ranged interactions and speculate on two ways in which unconventional superconductivity could be found in the 2DEG in physically realizable semiconductor heterostructures.  (We shall present more complicated examples in a forthcoming publication. \cite{RaghuII})

 In the first scenario, an in-plane magnetic field 
 is applied to partially polarize the 
 2DEG in a narrow quantum well.  This system 
 is predicted to exhibit p-wave pairing with a transition temperature which is non-monotonic in the magnetic field.  The optimal transition temperature is 
 obtained  for a magnetic 
field 
at which the ratio of the distinct spin Fermi momenta is $\eta=1/2$.  In the second scenario, 
the 2DEG is confined to a relatively broad quantum well, and the density is tuned to the range in which 
 two transverse subbands  
 are occupied.  
For fixed total electron density, the ratio, $\eta^2$, of densities in the two subbands increases with increasing thickness $w$ of the quantum well.  When this ratio is small, a p-wave 
groundstate arises, with a $T_c$ that rises sharply with increasing $\eta$ so long as $\eta < 1/2$.  However, this gives rise 
to a d-wave ground states above 
a certain critical thickness.

Insight into the dependence of $V/J$ on the thickness, $w$, is obtained by considering the 
Coulomb interactions.  A simple estimate shows that 
for $k_Fw \ll 1$, $V \sim e^2/ k_F$ and $ J \sim 
V\left( w k_F \right)$.  
Therefore, for thicker quantum wells, $J$ becomes 
increasingly important and favors d-wave pairing whereas thinner quantum wells should exhibit p-wave pairing.    
Depending on the ratio of $V/J$, the optimal $T_c$ occurs either for $\eta \approx 1/2$ (p-wave) or for the largest possible $\eta$ (d-wave).  In both cases, $T_c \sim E_F \exp{\left[ -\alpha/ (\rho V)^2 
 \right]}$ where $\alpha$ is an 
$\mathcal{O}(1)$ constant.  We have found that for d-wave superconductivity in the 2 subband system, 
values as low as  $\alpha \sim 1$ are within reach.

Three practical considerations warrant mention.  Due to the unconventional nature of the 
superconductvity, 
it is very fragile to even weak quenched disorder.  Therefore, the results  presented here are likely to be realized only in the purest samples with mean free paths exceeding  the Fermi wavelength by several orders of magnitude.  Furthermore, for small $r_s$, the plasma frequency is small compared to the Fermi energy, {\it i.e.} $\omega_p \sim \sqrt{r_s}E_F$, so even if it is reasonable to treat the interactions as short-ranged at low energies, this approximation is certainly not valid all the way to the Fermi energy.  Finally, since the transition temperatures  are exponentially low in the effective interactions, 
ultimately the superconductivity studied here is likely to be observable only in the regime $r_s \sim 1$, where the long-range character of the Coulomb interaction may not be negligible, and where, even for short-range interactions, a well-controlled solution to the problem is unfeasible.  
We therefore are forced rely on the hope that the asymptotic results  smoothly extrapolate  to the intermediate coupling regime, where it is conceivable that these states can be observed in experiment.

With these caveats, we turn to the most uncertain part of the discussion, and make the following {\it crude} quantitative estimate of $T_c$ based on our calculations:  We identify $V$ with the Fourier transform of the Coulomb interaction evaluated at $k_F$, {\it i.e.} $V  \approx e^2 \pi/k_F$, from which it follows that $\rho V \approx  (r_s/4)$.  Since we are going to extrapolate to $r_s \sim 1$ in any case, we simply ignore subtleties associated with the small value of $\omega_p$.  Then,
$T_c \sim E_F\exp[-\alpha (4/r_s)^2]$, where, for optimal circumstances $\alpha \approx 1$.

{\bf Acknowledgements} {  We would like to acknowledge important discussions with D. Scalapino in early stages of this project.  We thank  S. Das Sarma, E. Fradkin, D. Scalapino, and especially A. Chubukov for their insightful comments.  This work was supported, in part, by NSF grant number DMR-0758356 at Stanford.}

\appendix*
\section{Perturbation Theory}
In this section, we derive the perturbative expansion of the effective interaction in the Cooper channel for 
the problems studied in the main text.  Due to the presence of multiple bands, adopting a more compact notation enables us to treat all of the above problems in a unified fashion.  
We consider a Hamiltonian of the form 
\begin{eqnarray}
H&=& H_0 + H_1 \nonumber \\
H_0 &=&  \sum_{\bm k , a} \epsilon_{\bm k, a} c^{\dagger}_{\bm k, a} c_{\bm k, a} \nonumber \\
H_1 &=& \sum_{\bm k_1, \bm k_2, \bm k_3} \sum_{a,b,c,d}V_{ab,  cd} c^{\dagger}_{\bm k_1,  a} c^{\dagger}_{\bm k_2,  b} c_{\bm k_3,  c} c_{\bm k_4,  d}
\end{eqnarray}
where $\bm k_4 = \bm k_1 + \bm k_2 - \bm k_3$.  The Latin subscripts denote a collective set of ``band indices" which label the energy eigenstates.  In the problem of partially polarized Fermi surfaces, 
they simply label the majority and minority spin bands.  In the problem of multiple subbands in a quantum well, it indexes the subbands, and in the problem involving Rashba spin-orbit coupling, the Latin 
index refers to the positive and negative helicity subband.  
      The bare interaction $H_1$ is to be interpreted as a matrix; it's rows labels the outgoing states  and  its columns label the incoming states.  
\begin{figure}
\includegraphics[width=3.0in, angle = -90]{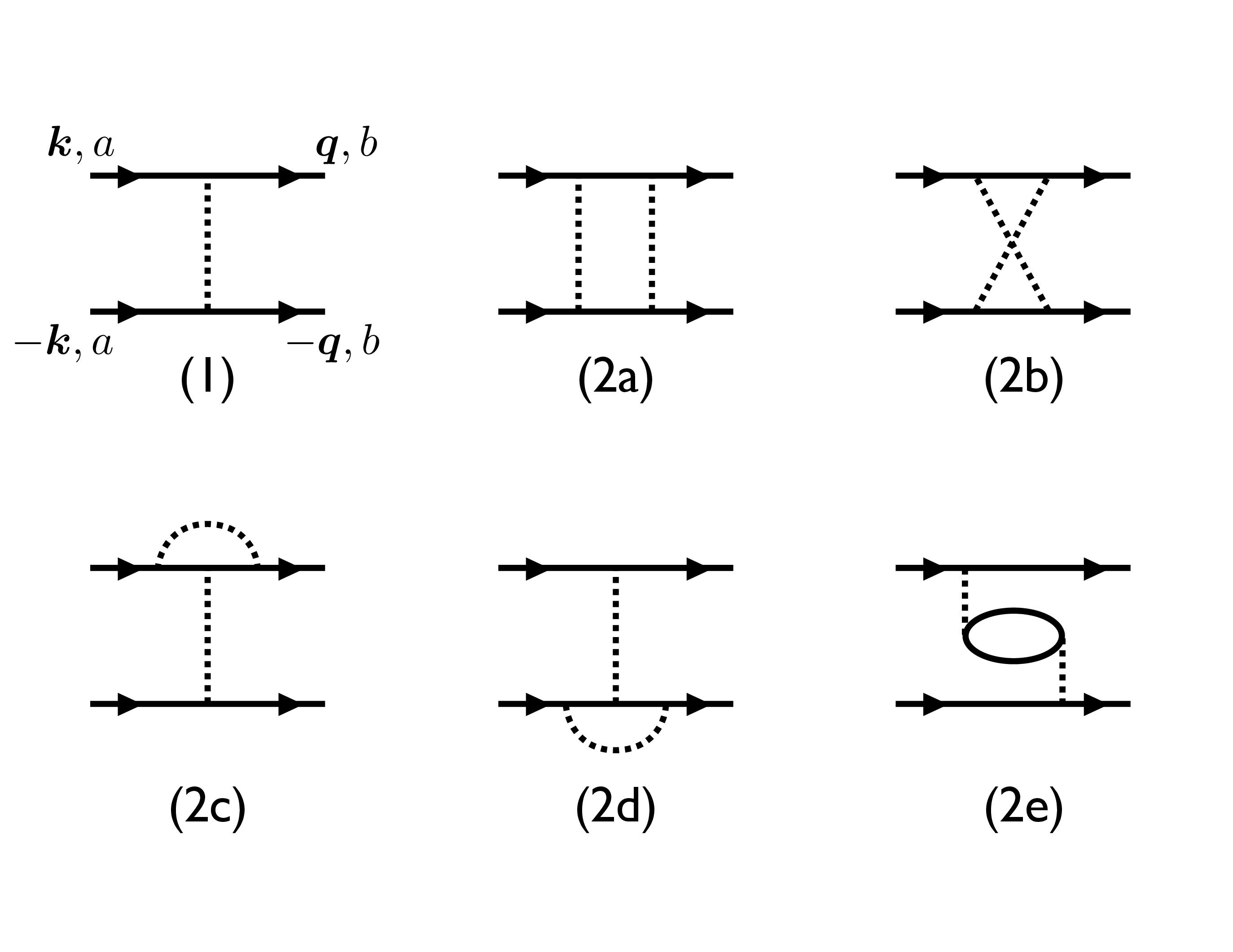}
\caption{ Diagrams which contribute to $V(\bm k, \bm q)$, shown to quadratic order in the interactions.  
Each of the incoming quasiparticles has momentum $\pm \bm k$ and band index $a$.  The outgoing electrons have momentum $\pm q$, and are in band $b$ (with the exception of the first diagram, the momenta and band indices of each diagram are not shown).  Both intra-band and inter-band scattering processes contribute to the effective interaction. }   
\label{diag}
\end{figure} 

Higher order scattering processes are derived using diagrammatic perturbation theory in the usual manner\cite{AGD}.  In addition to integrating over the internal momenta, the band indices of any internal line are also summed over, weighted by the appropriate component of the interaction vertex, as will be made clear from the examples below.   

The primary quantity of interest here is the two-particle scattering amplitude  in the Cooper channel, denoted $\Gamma(\bm k, \bm q)$, which is the amplitude for scattering a pair of electrons with momenta $\pm \bm  k$ into a pair with 
momenta $\pm \bm q$.  If the system at hand possesses inversion symmetry (so that the Kinetic energy consists of terms that are even powers of momentum), superconducting states can be classified has being even or odd parity states; the former include (e.g. s-wave, d-wave, etc.), and the latter include, (p-wave, f-wave, etc.)  instabilities are perfectly decoupled from one another.    If, in addition to inversion symmetry, spin-rotation symmetry is also preserved, then the scattering amplitudes in the singlet channel consist of processes in which the incoming electrons have opposite spin polarizations, whereas in the triplet channel, they have identical spin polarizations.  On the other hand, when inversion symmetry is broken by the presence of Rashba 
spin-orbit coupling, there is no sharp distinction between even and odd parity pairing.  However, 
since the Rashba coupling breaks the 2-fold degeneracy of single particle states at each momenta, 
there is pairing between states of opposite helicity (i.e. opposite momenta and opposite in-plane component of the spin).  

Figure \ref{diag} shows the lowest order Feynman diagrams which contribute to $\Gamma(\bm k, \bm q)$.  Generally, all of these diagrams contribute both in the singlet and triplet channel.  
The diagrams are each equivalent to
\begin{eqnarray}
&&1: \ \ \   V_{bb,aa} \nonumber \\
&& 2a: \ \ \  \sum_c  V_{bb,cc}V_{cc,aa}  \int_p G_c(-p)G_c(p) \nonumber \\
&& 2b: \ \ \   \sum_{c, d} V_{bc,ad} V_{db,ca} \int_p G_c(p)G_d(p+k+q) \nonumber \\
&& 2c: \ \ \   \sum_{c, d} V_{bc,ad} V_{db,ac} \int_p G_c(p)G_d(p+q-k) \nonumber \\
&& 2d: \ \ \   \sum_{c, d} V_{bd,ca} V_{bc,ad} \int_p G_c(p)G_d(p+k-q) \nonumber \\
&& 2e: \ \ \   -\sum_{c, d} V_{bd,ca} V_{cb,ad} \int_p G_c(p)G_d(p+k-q) \nonumber \\
\end{eqnarray}
where 
\begin{eqnarray}
\int_p \equiv \int \frac{ d \omega_p d^2 p}{\left( 2 \pi \right)^3} 
\end{eqnarray}
and 
\begin{equation}
G(p) = \frac{1}{ i \omega_p - \epsilon_{\bm p}}
\end{equation}
is the single particle Green function of the non-interacting system.  
We next apply this general formalism to the three problems studied in this paper.  
\subsection{Partially polarized Fermi surfaces}
Although this problem is rather simple, and the Feynman rules for a single band system 
are sufficient, we will apply the notation above to this problem.  This will certainly prove to 
be valuable in the more non-trivial examples studied thereafter.  
The interaction vertex, in the basis $( \uparrow \uparrow, \uparrow \downarrow, \downarrow \uparrow, \downarrow \downarrow)$, is
  \begin{equation}
V_{ab, cd} = 
\left( \begin{array}{cccc}
0 & 0 & 0 & 0 \\
0 & 0 & U & 0 \\
0 & U & 0 & 0 \\
0& 0 & 0 & 0
\end{array} \right).
\end{equation}
Note that in this problem, there are no processes which scatter two electrons from one Fermi 
surface, to two electrons in a different Fermi surface.  Therefore, $\Gamma(\bm k, \bm q)$ is 
diagonal in the band index - which is just a long-winded way of saying that only equal spin pairing (i.e. spin-triplet pairing) can occur.  
Indeed, the only diagram which contributes to the effective interaction is 2e, which yields:
\begin{eqnarray}
\Gamma_{\uparrow \uparrow}(\bm k, \bm q) &=& - U^2 \chi_{\downarrow}(\bm k - \bm q) + \mathcal{O}(\Omega_0) \nonumber \\
\Gamma_{\downarrow \downarrow}(\bm k, \bm q) &=& - U^2 \chi_{\uparrow}(\bm k - \bm q) + \mathcal{O}(\Omega_0)
\end{eqnarray}
where 
\begin{eqnarray}
\chi_{\sigma}(\bm k) &=& \int_p G_{\sigma}(p)G_{\sigma}(p+k) \nonumber \\
&=& - \int \frac{d^2 p}{(2 \pi)^2} 
\frac {f(\epsilon_{\bm p +\bm  k, \sigma}) - f(\epsilon_{\bm p, \sigma}) }{  \epsilon_{\bm p + \bm k, \sigma} - \epsilon_{\bm p, \sigma}}
\end{eqnarray}
is the non-interacting susceptibility of each spin band.  

\subsection{Multiple subbands in a 2DEG quantum well}
 For the problem involving two-subbands in a quantum well, we choose 
 the basis to be  $(1 \sigma 1 \sigma', 1 \sigma 2 \sigma', 2 \sigma 1 \sigma', 2 \sigma 2 \sigma')$ and 
\begin{equation}
V_{\alpha \beta,  \gamma \delta} = 
\left( \begin{array}{cccc}
U_{\sigma \sigma'} & 0 & 0 & J_{\sigma \sigma'} \\
0 & 0 & V_{\sigma \sigma'} & 0 \\
0 & V_{\sigma \sigma'} & 0 & 0 \\
J_{\sigma \sigma'} & 0 & 0 & U_{ \sigma \sigma'}
\end{array} \right)
\end{equation}
where
\begin{eqnarray}
U_{\sigma \sigma'} &=& U\left( 1 - \delta_{\sigma \sigma'} \right) \nonumber \\
J_{\sigma \sigma'} &=& J\left( 1 - \delta_{\sigma \sigma'} \right) \nonumber \\
V_{\sigma \sigma'} &=& V
\end{eqnarray}
In this basis $\sigma$ and $\sigma'$ refer to the spins of the incoming electron states.  
Since there is inversion and spin-rotational symmetry in this problem, we can study the 
effective interaction in the singlet ($\sigma = - \sigma'$) and triplet ($\sigma = \sigma'$) channels separately.   We shall refer to the effective interaction as $\Gamma_{s(t)}$ where the subscript stands for singlet and triplet.   Having specified the spin polarizations of the electrons, $\Gamma_{s(t)}$ will still be 
a $2 \times 2$ matrix due to the presence of two subbands.  We will denote this as 
\begin{equation}
\Gamma_{s(t)}(\bm k_a, \bm q_b) = \left( \begin{array}{cc} 
\Gamma_{s(t)}(\bm k_1, \bm q_1) & \Gamma_{s(t)}(\bm k_1, \bm q_2)\\
\Gamma_{s(t)}(\bm k_2, \bm q_1) & \Gamma_{s(t)}(\bm k_2, \bm q_2)
\end{array} \right)
\end{equation}
with the understanding that $\bm k_a$ denote momentum states associated with band a.  

Next, we state the contributions from each of the diagrams in Fig. \ref{diag}.  First, in the singlet channel, 
\begin{eqnarray}
\Gamma_s(1)&=& \left( \begin{array}{cc} U & J \\ J & U \end{array} \right) \nonumber \\
\Gamma_s(2a)&=&   \left( \begin{array}{cc} U^2 P_1 + J^2 P_2 & UJ(P_1 + P_2) \\ UJ(P_1 + P_2) & U^2 P_2 + J^2 P_1 \end{array} \right) \nonumber \\
\Gamma_s(2b)&=&   \left( \begin{array}{cc} U^2 \chi_{1,1}(\bm k_1 + \bm q_1) & VJ \chi_{1,2}(\bm k_1  + \bm q_2) \\ 
VJ \chi_{2,1}(\bm k_2  + \bm q_1) & U^2 \chi_{2,2}(\bm k_2 + \bm q_2) \end{array} \right) \nonumber \\
\Gamma_s(2c)&=&   \left( \begin{array}{cc} 0& VJ \chi_{1,2}(\bm k_1  - \bm q_2) \\ VJ\chi_{2,1}(\bm k_2  - \bm q_1) & 0 
\end{array} \right) \nonumber \\
\Gamma_s(2d)&=&   \left( \begin{array}{cc} 0& VJ \chi_{1,2}(\bm k_1  - \bm q_2) \\ VJ\chi_{2,1}(\bm k_2  - \bm q_1) & 0 
\end{array} \right) \nonumber \\
\Gamma_s(2e)&=&  - \left( \begin{array}{cc}  2V^2\chi_{2,2}(\bm k_1  - \bm q_1) &0 \\ 0 &  2V^2\chi_{1,1}(\bm k_2  - \bm q_2)
\end{array} \right) \nonumber \\
\end{eqnarray}
where 
\begin{eqnarray}
P_a &=& \int_p G_a(p) G_a(-p) \nonumber \\
&=& \rho_a \log\left[ A/ \Omega_0 \right] + \mathcal{O}(\Omega_0) \nonumber \\
\chi_{a,b}( \bm k) &=& \int_p G_a(p + k) G_b(p) \nonumber \\
&=& \int \frac{d^2 p}{(2 \pi)^2} \frac{f(\epsilon_{\bm p + \bm k, a}) - f(\epsilon_{\bm p, b})}{\epsilon_{\bm p + \bm k, a} - \epsilon_{\bm p, b}} + \mathcal{O}(\Omega_0) \nonumber \\
\end{eqnarray}
are the particle-particle and particle-hole susceptibilities, respectively.  

In the triplet channel, only diagram 2e has a contribution  and $\Gamma_t(\bm k_a, \bm q_b)$ is diagonal in subband index:
\begin{eqnarray}
\Gamma_t(\bm k_1, \bm q_1) &=& 
- U^2  \chi_{1,1}(\bm k_1 - \bm q_1) - 2 V^2 \chi_{2,2}(\bm k_1 - \bm q_1) \nonumber \\
\Gamma_t(\bm k_2, \bm q_2) & = &  -U^2 \chi_{2,2}(\bm k_2 - \bm q_2) - 2 V^2 \chi_{1,1}(\bm k_2 - \bm q_2) \nonumber \\
\Gamma_t(\bm k_1, \bm q_2) &=& 0
\end{eqnarray}
Having computed the effective interaction $\Gamma_{s,t}$, we define the quantity 
\begin{eqnarray}
g_{s,t}(\bm k_a, \bm q_b) \equiv \sqrt{ \frac{\bar v_{f}}{v_f (\hat k_a)}} \Gamma(\hat k_a, \hat q_b) 
\sqrt{ \frac{\bar v_{f}}{v_f (\hat q_b)}} 
\end{eqnarray}
which is also a matrix whose row and column indices are the set of momentum states on the Fermi surface.  

\subsection{Some relevant integrals}
In this section, we compute the particle-hole susceptibility matrix of the two subband problem:
\begin{equation}
\chi_{a,b}( \bm k) =  -\int \frac{d^2 p}{(2 \pi)^2} \frac{f(\epsilon_{\bm p + \bm k, a}) - f(\epsilon_{\bm p, b})}{\epsilon_{\bm p + \bm k, a} - \epsilon_{\bm p, b}} 
\end{equation}
We let
\begin{eqnarray}
\epsilon_{\bm k, 1} &=& \epsilon_{\bm k} \nonumber \\
\epsilon_{\bm k, 2} &=& \epsilon_{\bm k} + \Delta 
\end{eqnarray}
and set $\Delta = (k_{F1}^2 - k_{F2}^2)/2m > 0$ without loss of generality.  
It follows that the intraband susceptibilities are
\begin{equation}
\chi_{aa}(\bm q) =   2\int_0^{k_{Fa}} \frac{k d k }{(2 \pi)^2} \int_0^{2 \pi}  \frac{ d \theta}{ \epsilon_{\bm q} + k q \cos{\theta}/m}
\end{equation}
where $k_{F1} = \left(2 m \mu \right)^{1/2}$ and $k_{F2} = \left(2 m (\mu + \Delta) \right)^{1/2}$.  
The integrals are standard, resulting in  the following:
\begin{equation}
\chi_{a,a} = \frac{m}{2 \pi} \left[ 1-\frac{{\rm Re} \sqrt{q^2 - \left( 2 k_{Fa} \right)^2} } {q}  \right]
\end{equation}

\begin{figure}
\includegraphics[width=3.0in]{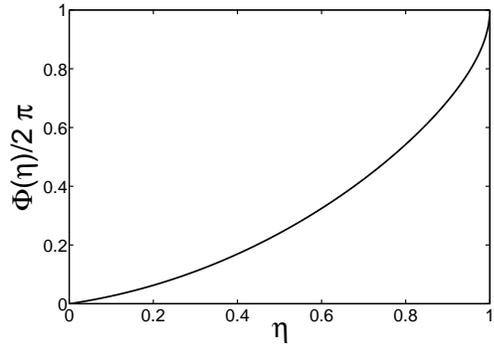}
\caption{ The scaling function $\Phi(\eta)$ which determines the effective interband interaction $\tilde g_{1,2}^{s,2}$.   }   
\label{interbandchi}
\end{figure}

The interband susceptibility is
\begin{eqnarray}
\chi_{1,2}(\bm q) = &&  \int_0^{k_{F1}} \frac{k d k }{(2 \pi)^2} \int_0^{2 \pi}  \frac{ d \theta}{ \epsilon_{\bm q} - \Delta + k q \cos{\theta}/m} \nonumber \\ 
&& \int_0^{k_{F2}} \frac{k d k }{(2 \pi)^2} \int_0^{2 \pi}  \frac{ d \theta}{ \epsilon_{\bm q} + \Delta + k q \cos{\theta}/m} \nonumber \\ 
\end{eqnarray}
These integrals are straightforward and they produce the final result: 
\begin{widetext}
\begin{equation}
\label{chi12}
\chi_{1,2}(q) = - \frac{\rho}{2} \left[ \frac{ {\rm Re} \sqrt{q^2\left[1 - \lambda(q) \right]^2 - \left(2 \eta k_{F 2} \right)^2 } }{q}   + \frac{ {\rm Re} \sqrt{q^2 \left[1 + \lambda(q) \right]^2 - \left(2 k_{F 2} \right)^2   } }{q}  
- \left[ 1 + \lambda(q) \right] - \left| 1 - \lambda(q)  \right|
\right]
\end{equation}
\end{widetext}
where 
\begin{eqnarray}
\eta&=& \frac{k_{F1}}{k_{F2}} \nonumber \\
\lambda(q) &=& \frac{k_{F2}^2 }{q^2} \left( 1 - \eta^2 \right) 
\end{eqnarray}
The effective inter-band attraction $\tilde g_{1,2}^{s,m}$ in the 
singlet channel is related to this susceptibility, as discussed in section 
\ref{2subband}:
\begin{eqnarray}
\label{g12}
\tilde g_{1,2}^{(s,m)} &=& 2VJ \rho \int \frac{d \theta}{2 \pi} \chi_{1,2}
\left( k_{\theta}   \right) \cos\left( m \theta \right)  \nonumber \\
k_{\theta} &=& k_{F2} \sqrt{\left(1-\eta \right)^2 + 4 \eta \sin^2 \left( \theta/2 \right)  } 
\end{eqnarray}
It is easy to show that the first two terms in Eq. \ref{chi12} do not contribute to $\chi_{1,2}(k_{\theta})$ 
since  
\begin{equation}
\frac{ \sqrt{k_{\theta}^2\left[1 - \lambda(k_{\theta}) \right]^2 - \left(2 \eta k_{F 2} \right)^2 } }{k_{\theta}} 
= \frac{2 \eta}{k_{\theta}} \sqrt{-\sin^2 \theta   }
\end{equation}
is purely imaginary for $0 \le \eta \le 1$.  Thus, 
\begin{equation}
\chi_{1,2}(k_{\theta}) = \frac{\rho}{2} \left[ 1 + \lambda(k_{\theta}) + \left | 1 - \lambda(k_{\theta}) \right| \right]
\end{equation}
This in turn can be rewritten as 
 \begin{eqnarray}
\chi_{1,2}(k_{\theta})= \rho \left\{\begin{array}{c c} 1 ,& \cos{\theta} < \eta \\  \frac{ \left(1 - \eta^2 \right)}{\ell_{\theta}^2} ,& 
\cos{\theta} > \eta  \end{array}\right.
\end{eqnarray}
where $\ell_{\theta} = k_{\theta}/k_{F2}$.    The integration in Eq. \ref{g12} is performed in the complex plane 
defining $z = -i \theta$.  
We find, for the d-wave case ($m=2$), 
\begin{equation}
 \tilde g_{1,2}^{(s,2)} = -\frac{V J \rho^2}{2 \pi}  \Phi(\eta)
 \end{equation}
where, for $0 \le x < 1$, 
\begin{equation}
\Phi(x) =  \frac{\pi x^4 + 2 \sin^{-1}x}{x^2} - 2 \frac{\sqrt{1-x^2}}{x}
\end{equation}
and $\Phi(1) = 0$.  The function $\Phi(x)$ is shown in Fig. \ref{interbandchi}.  
It should be noted  that $\Phi(x)$ is discontinuous at $x=1$, which results from the singluar behavior of the interband susceptibility in the  limit where the bands become degenerate: 
\begin{equation}
\lim_{\eta \rightarrow 1} \lim_{q \rightarrow 0} \chi_{1,2}(q) = \rho \frac{1+ \eta^2}{1- \eta^2}.
\end{equation}
However, this feature does not have a physical consequence since there is always a non-zero splitting 
between the bands caused by the finite thickness of the semiconductor heterostructure.

\bibliography{SmallU2deg}

\end{document}